\documentclass[prl,twocolumn,showpacs,floatfix,amsfonts]{revtex4}
\usepackage{graphicx,graphics,color,epsfig}
\usepackage{bm}
\usepackage{amsmath}
\usepackage{amssymb}
\usepackage{epstopdf}

\newcommand{\beq}{\begin{equation}}
\newcommand{\eeq}{\end{equation}}
\newcommand{\beqa}{\begin{eqnarray}}
\newcommand{\eeqa}{\end{eqnarray}}

\begin{document}

\title{Scalar spin chirality and  quantum Hall effect on triangular lattice}
\author{Ivar Martin}
\affiliation{Theoretical Division, Los Alamos National Laboratory, Los Alamos,
New Mexico 87545, USA}
\author{C. D. Batista}
\affiliation{Theoretical Division, Los Alamos National Laboratory, Los Alamos,
New Mexico 87545, USA}

\date{\today}
\begin{abstract}
We study the Kondo Lattice and Hubbard models on a triangular lattice for band filling factor 3/4.
We show that a simple non-coplanar chiral spin ordering (scalar spin chirality) is naturally realized in both models
due to perfect nesting of the Fermi surface. The resulting triple-${\bf Q}$ magnetic ordering is a natural counterpart
of the collinear Neel ordering of the half-filled square lattice Hubbard model. We show that
the obtained chiral phase exhibits a spontaneous quantum Hall-effect with $\sigma_{xy}= e^2/h$.
\end{abstract}
\maketitle

Interactions commonly lead to magnetism. For instance, the ground state of the half-filled Hubbard model becomes 
antiferromagnetic (AFM) for a finite onsite Coulomb interaction U. The AFM phase is collinear on bipartite lattices 
and it exists for any finite $U/t$ ($t$ being the nearest-neighbor hopping) in the cases of square and cubic lattices 
due to nesting properties of the Fermi surface (FS). The situation is less clear when competing interactions or lattice 
geometry lead to frustration and more complex magnetic orderings. For instance, a $120^{\circ}$ three-sublattice 
(non-collinear) structure is stabilized on the half-filled triangular lattice for large enough $U/t$ and a spin 
liquid ground state has been proposed for the kagome lattice. The {\em non-coplanar} spin orderings have proven to 
be particularly elusive.  Besides being the next level in magnetic ordering complexity, 
the non-coplanar states have unusual transport~\cite{Ohgushi} and magneto-electric \cite{Lev08} properties resulting from their 
chiral structure. Consequently, it is important to find simple and realistic models in which such 
states could emerge and remain stable over a range of Hamiltonian parameters.

A triangular plaquette is a basic building block for lattices with built-in geometrical frustration.  Assume that 
on every site $i$ there is an ordered magnetic moment ${\bf S}_i$.  If the moments on the plaquette are non-coplanar, 
the resulting {\rm scalar spin chirality}, $\langle {\bf S}_i \cdot {\bf S}_j \times {\bf S}_k\rangle \neq 0$,
breaks both the time-reversal and the parity symmetries.  Notably, the scalar spin chirality can exist even without 
having statically-ordered magnetic moments ($\langle {\bf S}_i\rangle=0$) \cite{Domenge05}.
When conduction electrons propagate through such a spin texture,  they may exhibit spontaneous
Hall effect in the absence of any externally applied magnetic field \cite{Taguchi01}.   Moreover, if the spin ordering 
opens a full gap in the charge excitation spectrum,  a spontaneous  {\em quantum} 
Hall insulator can form \cite{Ohgushi}.  This exotic phenomenon has the following origin: Due to the exchange interaction, 
the underlying local moment texture aligns the spins of the conduction electrons
inducing a Berry phase in the electron wavefunction, which is indistinguishable from a ``real" magnetic flux.  
Ohgushi {\em et al.}
\cite{Ohgushi}, considered a model of 2D kagome ferromagnet where spin orbit
interaction tilts the local moments so as to produce non-zero chirality.  The
spin configuration corresponded to total zero fictitious flux through the unit
cell, while still creating non-zero quantized Hall conductivity at filling 1/3.  The authors were primarily concerned with  
the double exchange model, which assumes two species of electrons -- the localized ones and the itinerant ones, 
but also suggested that similar physics may occur in a single band Hubbard-like model in the presence 
of spin-obit interaction.  Appearance of quantum Hall effects in systems with spontaneous {\em  bond} ordering has recently been discussed in Raghu {\em et al.} \cite{Zhang}.

\begin{figure}[ht]
\includegraphics[width=.9\columnwidth]{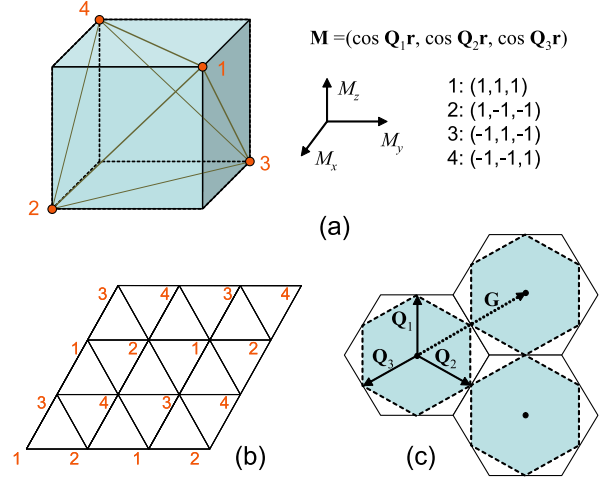}
\caption[]{(Color online) Chiral spin ordering on a triangular lattice.  (a)  Four possible orientations of the local
magnetic moments.  They correspond to the normals to the sides of a regular tetrahedron.
(b) Arrangement of the local moments on the real space lattice. ${\bf a}_1$ and ${\bf a}_2$ are the
lattice vectors.  (c) Momentum space.  Unshaded hexagons define the Brillouin zones.  $\bf G$ is one of the
reciprocal lattice vectors.   Vectors ${\bf Q}_i$ are the ordering vectors for the ``tetrahedral"
chiral order ($Q_i = G/2$).  Note that ${\bf Q}_i$ perfectly nest the 3/4-filled FS (shaded hexagons).}
\label{FIG:1}
\end{figure}

In the present work we consider a model of itinerant electrons interacting with a 
chiral magnetic state on a simple triangular lattice.   The magnetic ordering has the four-site unit cell, 
with the local spin orientations corresponding to the normals to the faces of a regular tetrahedron, see Fig.~\ref{FIG:1}.  
This magnetic structure has been earlier proposed as a likely candidate for a low temperature phase of Mn monolayer on 
Cu(111) surfaces \cite{Kurz} and a possible nuclear spin ground state of a two-dimensional $^3$He \cite{Momoi}.  
We demonstrate that in the presence of itinerant electrons, this structure naturally appears as a weak-coupling 
instability at 3/4 filling due to the nesting properties of the electronic Fermi surface, in direct 
analogy with the magnetic instability in the square-lattice Hubbard model at half filling.  
In the insulating state at 3/4 filling the model exhibits spontaneous quantum Hall effect.

Let us first consider the Kondo-Lattice Hamiltonian,
\begin{equation}
H = -t \sum_{\langle i j \rangle}{c_{i\alpha}^\dagger c_{j\alpha}} -
\frac{J}{2} \sum_i{{\bf S}_i \cdot c_{i\alpha}^\dagger{\bm
\sigma}_{\alpha\beta} c_{i\beta}},
\label{eq:Hde}
\end{equation}
which describes electrons hopping between the nearest-neighbor sites of
triangular lattice interacting via on-site exchange coupling $J$ with the localized moments ${\bf S}_i$. 
The operator $c^\dagger_{i\alpha}$
($c^\dagger_{i\alpha}$) creates (destroys) an electron with spin $\alpha$ on
site $i$ and ${\bm \sigma}_{\alpha\beta} = (\sigma^x_{\alpha\beta},
\sigma^y_{\alpha\beta}, \sigma^z_{\alpha\beta})$ is  a vector of the Pauli
matrices.  This model may describe various $d$ and $f$-electron compounds, with $J$ being the ferromagnetic Hund's coupling in the first case and the antiferromagnetic Kondo exchange
in the latter.
We consider the classical limit $S \gg 1$ for the localized spins.  Then, the magnitude of $|{\bf S}_i| =  S$ is fixed and the only remaining degree of freedom is the spin orientation.  The sign of $J$ is irrelevant (the spectrum does not depend on this sign and the eigenstates are connected by a global time reversal transformation of the localized spins).  In the following, we take $J >0$.

The simplest limit of $H$ is the strong coupling, $J S/t \gg 1$.   Two disconnected sectors appear in this regime:
the low (high) energy subspace of states in which conduction electrons hop between neighboring
sites remaining parallel (antiparallel) to the local moment ${\bf S}_i$.
Within each sector, the conduction electron spin orientation is enslaved to its position, and the effective hopping amplitude acquires a quantum mechanical phase \cite{Anderson}.  This Berry phase is indistinguishable from the Aharonov-Bohm phase in an externally applied magnetic field \cite{Peierls}.  The effective Berry flux through a closed path is
half of the solid angle subtended by the electron spin while moving 
along the path \cite{Fradkin}.  The fluxes experienced by electrons with spins aligned  and anti-aligned 
with the local moments are equal in magnitude but opposite in sign.
For the specific case of ``tetrahedral" ordering, Fig.~\ref{FIG:1}, the solid angle subtended by local 
moments in corners of each elementary triangular plaquette is clearly $\Omega = 4\pi/4$.  Thus,
in the limit of strong coupling, the problem is equivalent to one of
electrons on a triangular lattice in a uniform spin-dependent magnetic field corresponding to flux $\phi_\sigma = \sigma \pi/2$ 
per triangular plaquette.  The overall energy splitting between the spin up and spin down sectors is $2JS$. The energy spectrum can be easily found,
\beq
\epsilon_{\bf k} = \pm JS \pm \frac{2t}{\sqrt3}\left(\cos^2{\bf k a}_1 + \cos^2{\bf k a}_2 + \cos^2{\bf k a}_3\right)^{1/2}\label{eq:E1}
\eeq
where ${\bf a}_1$ and ${\bf a}_2$ real space lattice vectors and ${\bf a}_3 =
{\bf a}_1 - {\bf a}_2$.  Since the real space (magnetic) unit cell has doubled, this dispersion is 
defined on {\em half} of the original reciprocal
unit cell. The amplitude of the hopping is reduced by a factor $\sqrt 3$ due to the angle $\theta$ 
between the neighboring spins, $t\rightarrow t \cos(\theta/2) = t/\sqrt3$.  There are total four 
energy-split bands -- (2 spins) $\times$ (2 sites in magnetic unit cell).  Thus for fillings that are 
integer multiples of 1/4, the system is an insulator.  
Filling 1/2 corresponds to full filling of the spin-up sector, and therefore it is equivalent to a band insulator, 
with $\sigma_{xy} =0$.  The ``interesting" fillings are 1/4 and 3/4.
From the explicit treatment of spinless electrons on triangular lattice \cite{HofstTri}, one finds that 
$|\sigma_{xy}| = e^2/{h}$ in both cases, but with opposite signs, since spin up and spin down electrons 
experience opposite effective magnetic fluxes.

Now we turn to the case of an arbitrary $JS/t$.  From Fig.~\ref{FIG:1}, the unit cell in the ordered state is {\em quadrupled} -- it is now spanned by vectors $2{\bf a}_1$ and $2{\bf a}_2$.  It is convenient to treat to problem in the momentum space.  For that we notice that the ``tetrahedral" magnetic ordering is equivalent to the so-called $3q$ non-coplanar state, with the order parameter
\begin{equation}
{\bf S}({\bf r}) = ({\cal S}_a \cos {\bf Q}_a {\bf r}, {\cal S}_b \cos {\bf Q}_b {\bf r},
{\cal S}_c \cos {\bf Q}_c {\bf r}). \label{eq:order}
\end{equation}
The amplitudes ${\cal S}_\alpha$ are equal in magnitude but can have different signs 
(arbitrary global rotation of the triad in the spin space yields an equivalent state).  
The ordering vectors ${\bf Q}_\alpha$ are equal to half of the reciprocal lattice vectors, see Fig.~\ref{FIG:1}c.  
Therefore, momenta $+{\bf Q}_{\eta} $ and $-{\bf Q}_{\eta} $ are equivalent (${\bf Q}_{\eta}
\cong -{\bf Q}_{\eta} $), and \mbox{$\pm{\bf Q}_a \pm{\bf Q}_b \pm{\bf Q}_c
\cong 0 $}.
The scalar chirality, defined as a counter-clock-wise triple product 
${\bf S}_1 \cdot({\bf S}_2 \times {\bf S}_3)\propto {\cal S}_a {\cal S}_b {\cal S}_c$ around a 
plaquette (e.g. 123 in Fig.~\ref{FIG:1}b) can be either positive or negative, depending on the 
choice of $S_\alpha$.  To be specific, let's chose all ${\cal S}_\alpha > 0$, Fig.~\ref{FIG:1}a.
In momentum space the Hamiltonian Eq.~(\ref{eq:Hde}) becomes
\begin{equation}
H = \sum_{{\bf k}\sigma} \epsilon_{\bf k}  c^\dagger_{{\bf k}\sigma} c^{\;}_{{\bf k}\sigma} -
J \sum_{{\bf k}\alpha \beta \eta} {\cal S}_{\eta} c^\dagger_{{\bf k}\alpha} \sigma^{\eta}_{\alpha \beta}
c^{\;}_{{\bf k +Q_{\eta}}\beta},
\end{equation}
where $\eta={a,b,c}$.
The momentum summation is over the quarter of the original reciprocal unit cell.  The Hamiltonian splits into two disjoint parts
\beqa
H &=& {\bf c}_I^\dagger\hat H{\bf c}_I + {\bf c}_{II}^\dagger\hat H{\bf c}_{II}\label{eq:He}\\
\hat H &=& \left(\begin{array}{cccc}\epsilon_{\bf k} & J{\cal S}_a & iJ{\cal S}_b & J{\cal S}_c \\
{\cal S}_a & \epsilon_{{\bf k + Q}_1} & -J{\cal S}_c & -iJ{\cal S}_b \\
-iJ{\cal S}_b & -J{\cal S}_c & \epsilon_{{\bf k + Q}_2} & J{\cal S}_a \\J{\cal S}_c & iJ{\cal S}_b & J{\cal S}_a &
\epsilon_{{\bf k + Q}_3}\end{array}\right)
\eeqa
where ${\bf c}_I^\dagger = (c^\dagger_{{\bf k}\uparrow} , c^\dagger_{{\bf k +
Q}_1\downarrow} ,c^\dagger_{{\bf k + Q}_2\downarrow},c^\dagger_{{\bf k +
Q}_3\uparrow})$ and ${\bf c}_{II}^\dagger = (c^\dagger_{{\bf k}\downarrow} ,
c^\dagger_{{\bf k + Q}_1\uparrow} ,-c^\dagger_{{\bf k +
Q}_2\uparrow},-c^\dagger_{{\bf k + Q}_3\downarrow})$.  Therefore, the spectrum
is (at least) doubly degenerate.  The physical reason for the degeneracy is the same 
as for the degeneracy between the spin-up and spin-down band in a mean
field treatment of an antiferromagnet on a square lattice -- the spin-up and
spin-down wave functions are related by a lattice translation combined with a
spin rotation. Indeed,  from Fig.~\ref{FIG:1}b, the local moment arrangement
changes as $(1\leftrightarrow2, \ 3\leftrightarrow4)$ under
translation by ${\bf a}_1$. From Fig.~\ref{FIG:1}a, a
spin rotation by angle $\pi$ around $x$ axis restores the initial state. 
Therefore, the composition of both transformations leaves $\hat H$ unchanged, 
while $c_{II}\leftrightarrow c_I$.

\begin{figure}[ht]
\includegraphics[width=.9\columnwidth]{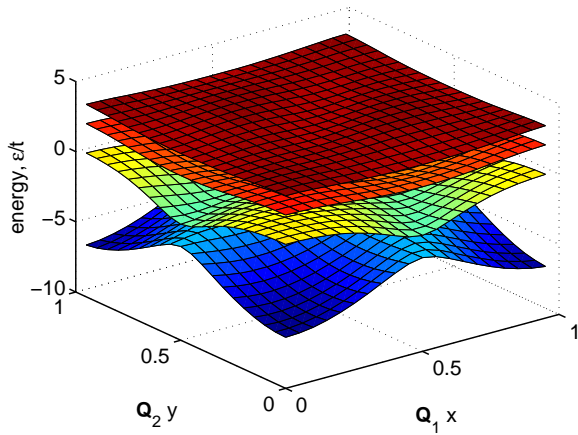}
\caption[]{(Color) Band structure corresponding to Eq.~(\ref{eq:He}).  The coupling strength is $JS/t = 1$.
The axes are chosen along ${\bf Q}_a$ and ${\bf Q}_b$ and point $(0,0)$ corresponds to the center of 
the first Brillouin zone.}
\label{FIG:bands}
\end{figure}

Fig.~\ref{FIG:bands} shows the band structure corresponding to $H$ in the reduced (quarter) reciprocal 
unit cell with the axes chosen along ${\bf Q}_a$ and ${\bf Q}_b$.  For an arbitrary value of $JS/t$, 
there are four pairs of degenerate bands.  For large enough $JS/t$, there is a gap between all pairs, and 
the dispersion is well described by Eq.~(\ref{eq:E1}).  The Hall conductivity can be calculated directly 
using the Kubo formula,
\beq
\sigma_{xy} = \frac{e^2}{h}\frac{1}{2 \pi i} \sum_{nm{\bf k}}{\frac {j_x^{nm} j_y^{mn} }{(\epsilon_{n \bf k} - \epsilon_{m \bf k})^2}[n_F(\epsilon_{n \bf k})-n_F(\epsilon_{m \bf k})]},\label{eq:sxy}
\eeq
with summation performed over all 8 bands $n,m$ and $(j_x,j_y)$ being the $x$ and $y$ components of the current 
operator.  At zero temperature, when the Fermi distribution 
functions $n_F$ are 1 for occupied bands and 0 for the empty ones, we recover the results for $\sigma_{xy}$ 
obtained earlier for $JS/t \gg 1$.  Each pair of degenerate bands contributes Chern number $\pm 1$, instead of  
$0$ or $\pm 2$, due to the additional local (in $\bf k$-space) $SU(2)$ symmetry implied by the degeneracy \cite{Kohmoto}.

For small $JS/t$, the lower three pairs of bands overlap.  However,
the gap between the two upper bands persists for arbitrarily small $JS/t$.  This signifies a weak coupling instability towards 3$q$ ordering at $3/4$ filling (1.5 electrons per site).  Indeed, at this filling the FS of non-interacting electrons is a regular hexagon inscribed into
the first Brillouin zone (see Fig.~\ref{FIG:1}c).  The vectors ${\bf Q}_{\eta}$ perfectly nest all sides of 
the FS, and hence a local interaction will cause the $3q$ instability. In general, an ordering 
${\cal \vec S}_\alpha \cos({\bf Q}_\alpha r)$ with arbitrarily directed ${\cal \vec S}_\alpha$ is expected.  
However, the constraint of same magnetization magnitude on every lattice site $|{\bf S}_i|=S$ requires that 
${\cal \vec S}_\alpha$ be mutually orthogonal, leading to Eq.~(\ref{eq:order}).  The quantum Hall conductivity 
at 3/4 filling for arbitrarily small $JS/t$ is $\sigma_{xy} = e^2/h$ at zero temperature.

\begin{figure}[ht]
\includegraphics[width=.9\columnwidth]{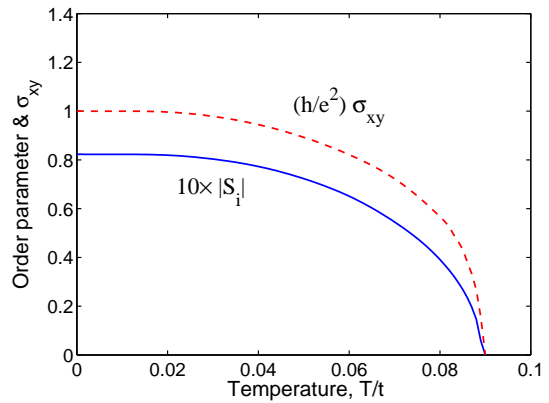}
\caption[]{Hubbard model, U = 4t. The magnitude of the order parameter for non-coplanar 3$q$ structure and Hall conductivity as a function
of temperature. }
\label{FIG:OP}
\end{figure}

We now consider the possibility that the local moments do not originate from localized 
electrons on different orbitals, but are generated self-consistently in a  single-band Hubbard model $H_{\rm Hubb}$
on a triangular lattice. The connection between this case and the one studied above is evident from the mean-field 
expression of $H_{\rm Hubb}$ \cite{Fradkin}:
\beq
H_{\rm Hubb} = -t \sum_{\langle i j \rangle}{c_{i\alpha}^\dagger c_{j\alpha}} -
\frac{2U}{3} \sum_i{2\langle{\bf S}_i\rangle{\bf S}_i - \langle{\bf S}_i\rangle^2}. \label{eq:Hmf}
\eeq
Now all the spin degrees of freedom are provided by the same (itinerant) electrons and 
$\langle {\bf S}_i \rangle = \langle c_{i\alpha}^\dagger{\bm \sigma}_{\alpha\beta} c_{i\beta} \rangle$.  The result of the self-consistent calculation for the non-coplanar 
$3q$ order parameter as a function of temperature is presented in Figure \ref{FIG:OP}.  The finite-temperature 
calculation of the transverse conductivity based on Eq.~(\ref{eq:sxy}) reveals that $\sigma_{xy}$ is a good proxy of the order parameter.

$H_{\rm Hubb}$ has the same form as Eq.~(\ref{eq:Hde}) but $\langle {\bf S}_i \rangle^2$ is no longer fixed
due to quantum fluctuations (${\bf S}_i$ is now in the $S=1/2$ quantum limit).  The relaxation of this constraint 
allows for more general $3q$ structures such as collinear or coplanar. The energy difference between the $3q$ collinear, coplanar,
and non-coplanar structures is very small, since it is caused by the interaction between the order parameters ${\cal \vec S}_\alpha$ 
at the corners of the FS.  Numerically we find that the coplanar (but non-collinear) phase is not favored, while the non-coplanar  and collinear \cite{Balents} $3q$ structures are are very close  in energy at the mean field level.  An analysis beyond mean-field would be required to determine the relative stability of the two phases.

A spontaneous breaking of a continuous symmetry is impossible at finite temperature in 2D for short-range interactions \cite{MW}.  
Does this imply that 
finite-temperature spontaneous quantum Hall effect is impossible?  To answer this question we need to analyze the symmetry of 
the problem in more detail. To that end we consider the Ginzburg-Landau theory for the $3q$ magnetic structure.  
It can be constructed in close analogy to the theory for the Heisenberg antiferromagnet (AF) on a square lattice.
However, instead of a single mean-field order parameter at momentum $(\pi,\pi)$, now there are three order parameters, 
with the ordering wavevectors ${\bf Q}_\alpha$.
To the lowest order, they do not interact and each is described by the standard AF Lagrangian,
${\cal L}_\alpha = [r + \omega^2 - ({\bf q}-{\bf Q}_\alpha)^2]|{\bf S}^\alpha_{\bf q}|^2 + u |{\bf S}^\alpha_{\bf q}|^4/2$, 
with appropriately scaled frequencies and momenta \cite{note}.
The transition is controlled by the parameter $r$ which can be tuned by, e.g. temperature.
Given the momentum space structure of the order parameters, the first symmetry-allowed term that selects
the mutual orientation of the order parameters is ${\cal L}_{int} = v [{\bf S}_1\cdot ({\bf S}_2\times{\bf S}_3)]^2$.  
Negative value of $v$ favors non-coplanar configuration with ${\bf S}_\alpha$ orthogonal to each other, while $v>0$ 
favors configuration with ${\bf S}_i$ collinear \cite{Balents}.  This term reduces the symmetry of the Lagrangian 
down to $SO(3)\times Z_2$; however, being higher order in $S$, it does not modify the mean-field transition temperature.  
Non-coplanar ordering corresponds to complete breaking of this symmetry.  Thus, by the number of $SO(3)$ generators, 
there are three gapless (linearly dispersing) Goldstone modes.  In addition, there are three modes that have a small 
gap $\sqrt{- v r^2/u^2}$ induced by the orientation-locking term.

The gapless modes are responsible for the destruction of the long-range magnetic ordering at finite temperature.  
However, smooth distortion of the magnetic texture does not necessarily destroy the chirality \cite{Vilain}.  
The chirality is a discrete (Ising) order parameter, which can be destroyed in two ways: Through point defects -- vortices, 
and line defects -- domain walls.  Since the order parameter space (the ``vacuum manifold") of the model is given by $SO(3)$, 
there are only two topologically distinct kinds of vortices -- the trivial and the non-trivial one \cite{Kawamura84}, in contrast 
to the more familiar $U(1)$ vortices, which can have an arbitrary integer winding number.
At the vortex core, one or more of the order parameters ${\bf S}_\alpha$ is completely suppressed.  The vortices effectively 
reduce the {\em magnitude} of average chirality, without changing its sign.  On the other hand, a domain wall 
corresponds to a line where the sign of chirality changes via mutual reorientation of the order parameters, 
without necessarily changing their magnitude (at the domain wall the state is coplanar).  
The finite temperature transition can be governed by either proliferation of domain walls \cite{Sachdev} or vortices \cite{Domenge08}.  
Which one is realized, depends on the relative importance of the individual order parameter 
stiffness and the chiral locking term in the free energy.  In the weak coupling Hubbard model 
it is likely that the chiral term is small, and thus the domain walls have lower energy cost.  Then, 
one may expect that the finite-temperature transition is in the generalized Ising class \cite{Momoi}.

In conclusion, we studied a model of itinerant electrons interacting with a chiral magnetic texture on a triangular lattice.  
We found that at 3/4 filling the system can become a spontaneous quantum Hall insulator, even without static magnetic ordering.
One possible material realization is Na$_{0.5}$CoO$_2$~\cite{Imai}.  

{\bf Acknowledgments:} We acknowledge useful discussions with D. Arovas, A. Auerbach, L. Boulaevskii, and N. Read.  This work
was carried out under the auspices of the National Nuclear Security
Administration of the U.S. Department of Energy at Los Alamos National
Laboratory under Contract No. DE-AC52-06NA25396 and supported by the LANL/LDRD
Program.

\end{document}